\documentclass[aps,prb,twocolumn,superscriptaddress]{revtex4-1}
\usepackage{amsfonts}
\usepackage{graphicx}
\usepackage{epstopdf}
\usepackage{epsfig}
\usepackage{amsmath}
\usepackage{bm}
\usepackage{color}
\usepackage{makecell}
\usepackage{array}
\usepackage{booktabs}
\usepackage{tabularx}

\begin{document}

\title{Chaos based Berry phase detector}

\author{Cheng-Zhen Wang}
\affiliation{School of Electrical, Computer and Energy Engineering, Arizona State University, Tempe, Arizona 85287, USA}

\author{Chen-Di Han}
\affiliation{School of Electrical, Computer and Energy Engineering, Arizona State University, Tempe, Arizona 85287, USA}

\author{Hong-Ya Xu}
\affiliation{School of Electrical, Computer and Energy Engineering, Arizona State University, Tempe, Arizona 85287, USA}

\author{Ying-Cheng Lai} \email{Ying-Cheng.Lai@asu.edu}
\affiliation{School of Electrical, Computer and Energy Engineering, Arizona State University, Tempe, Arizona 85287, USA}
\affiliation{Department of Physics, Arizona State University, Tempe, Arizona 85287, USA}

\date{\today}
\begin{abstract}

The geometric or Berry phase, a characteristic of quasiparticles, is
fundamental to the underlying quantum materials. The discoveries of new
materials at a rapid pace nowadays call for efficient detection of the
Berry phase. Utilizing $\alpha$-T$_3$ lattice as a paradigm,
we find that, in the Dirac electron optics regime, the semiclassical
decay of the quasiparticles from a chaotic cavity can be effectively
exploited for detecting the Berry phase. In particular, we demonstrate
a one-to-one correspondence between the exponential decay rate and the
geometric phase for the entire family of $\alpha$-T$_3$ materials. This
chaos based detection scheme represents an experimentally feasible way to
assess the Berry phase and to distinguish the quasiparticles.

\end{abstract}
\maketitle

\section{Introduction} \label{sec:intro}

The geometric phase, commonly referred to as the Pancharatnam-Berry phase or
simply the Berry phase, is a fundamental characteristic of the quasiparticles
of the underlying quantum material. When a system is subject to a cyclic
adiabatic process, after the cycle is completed, the quantum state returns
to its initial state except for a phase difference - the Berry
phase~\cite{Pancharatnam:1956,LHOPS:1958,Berry:1984}. In general, the exact
value of the Berry phase depends on the nature of the quasiparticles and
hence the underlying material. For example, the Berry phases in monolayer
graphene~\cite{ZTSK:2005,CU:2008} and graphite bilayers~\cite{MS:2008} are
$\pm\pi$ and $2\pi$, respectively. In $\alpha$-T$_3$ lattices, for different
values of $\alpha$, the Berry phases associated with the quasiparticles
are distinct~\cite{ICN:2015}.

Advances in physics, chemistry, materials science and engineering have led to
the discoveries of new materials at an extremely rapid pace, e.g., the various
two-dimensional Dirac materials~\cite{GN:2007,GG:2013,AKB:2016}. These
materials host a variety of quasiparticles with distinct physical
characteristics including the Berry phase. To be able to detect Berry
phase for a new material would generate insights into its physical properties
for potential applications. Conventionally, this can be done using the
principle of Aharonov-Bohm interference. For example, an atomic
interferometer was realized in an optical lattice to directly measure the
Berry flux in momentum space~\cite{DLRBSS:2015}. Graphene resonators subject
to an external magnetic field can be used to detect the Berry
phase~\cite{RL:2016,Ghaharietal:2017}. Specifically, for a circular graphene
$p$-$n$ junction resonator, as a result of the emergence of the $\pi$ Berry
phase of the quasiparticles (Dirac fermions) when the strength of the magnetic
field has reached a small critical value, a sudden and large increase in the
energy associated with the angular-momentum states can be detected. In
photonic crystals, a method was proposed to detect the pseudospin-1/2 Berry
phase associated with the Dirac spectrum~\cite{SNB:2008}.
In such a system, the geometric Berry phase acquired upon rotation of the
pseudospin is typically obscured by a large and unspecified dynamical phase.
It was demonstrated~\cite{SNB:2008} that the analogy between a photonic
crystal and graphene can be exploited to eliminate the dynamical phase, where
a minimum in the transmission arises as a direct consequence of the Berry
phase shift of $\pi$ acquired by a complete rotation of the pseudospin
about a perpendicular axis.

In this paper, we report a striking phenomenon in 2D Dirac materials, which
leads to the principle of chaos based detection of Berry phase. To be concrete,
we consider the entire $\alpha$-T$_3$ material family. An $\alpha$-T$_3$
material can be synthesized by altering the honeycomb lattice of graphene
to include an additional atom at the center of each hexagon which, for
$\alpha = 1$, leads to a $T_3$ or a dice lattice that hosts pseudospin-1
quasiparticles with a conical intersection of triple degeneracy in the
underlying energy band~\cite{Sutherland:1986,BUGH:2009,SSWX:2010,GSC:2010,
DKM:2011,WR:2011,HLHZC:2011,MWCZ:2012,MYAKBV:2013,Guzmanetal:2014,RPG:2015,
GCvdBO:2015,Lietal:2015,MSCGOAT:2015,VCMRMWSM:2015,TOINNT:2015,FZLC:2016,
DLKDD:2016,Zhuetal:2016,BCWVFCB:2016,FS:2017,Ezawa:2017,Zhongetal:2017,
ZZYXZ:2017,DOHL:2017,Slotetal:2017,Tanetal:2018}. An $\alpha$-T$_3$ lattice is
essentially an interpolation between the honeycomb lattice of graphene and
a dice lattice, where the normalized coupling strength $\alpha$ between the
hexagon and the central site varies between zero and
one~\cite{RMFPM:2014,PFRM:2015,ICN:2015,IN:2016,IN:2017,KDDC:2017}, as shown
in Fig.~\ref{fig:alpha-T3-cavity}(a). Theoretically, pseudospin-1
quasiparticles are described by the Dirac-Weyl
equation~\cite{BUGH:2009,SSWX:2010,BCWVFCB:2016}. Suppose we apply an
appropriate gate voltage to generate an external electrostatic potential
confinement or cavity of $\alpha$-T$_3$ lattice. The mechanism for Berry phase
detection arises in the short wavelength or semiclassical regime, where the
classical dynamics are relevant and can be treated according to ray optics
with reflection and transmission laws determined by Klein tunneling - the
theme of the emergent field of Dirac electron optics
(DEO)~\cite{CPP:2007,CFA:2007,DSBA:2008,
SRL:2008,BSAT:2009,MZ:2010,GRL:2011,WLLM:2011,RMLWRS:2013,LZEC:2013,HBF:2013,
AU:2013,WF:2014,Zhaoetal:2015,Rickhausetal:2015,Leeetal:2015,RMRS:2015,
WH:2016,CCOWK:2016,GBKPP:2016,Leeetal:2016,Chenetal:2016,SPBJ:2016,
LGR:2017,VHSWTG:2017,Jiangetal:2017,Ghaharietal:2017,ZZYC:2017,BCSCPB:2017,
XWHL:2018}. If the shape of the cavity is highly symmetric, e.g., a circle,
the classical dynamics of the quasiparticles are integrable. However, if the
cavity boundaries are deformed from the integrable shape, chaos can arise. We
focus on the energy regime $V_0/2<E<V_0$ in which Klein tunneling is enabled,
where $V_0$ is the height of the potential [Fig.~\ref{fig:alpha-T3-cavity}(b)],
so that the relative effective refractive index $n$ inside the cavity falls in
the range $[-\infty,-1]$. As a result, there exists a critical angle for total
internal reflections. For different values of the material parameter $\alpha$,
the physical characteristics of the quasiparticles, in particular the values of
the Berry phase, are different. Our central idea is then that, for a fixed
cavity shape, the semiclassical decay laws for quasiparticles corresponding
to different values of $\alpha$ would be distinct. If the classical cavity
dynamics contain a regular component, the decay laws will be
algebraic~\cite{MO:1985,LDGB:1992,HKL:2000,WHK:2002,CK:2008},
but we find that the differences among them will not be statistically
significant enough to allow lattices of different values of $\alpha$ to be
distinguished. However, when the cavity is deformed so that the classical
dynamics are fully chaotic, the decay law becomes exponential~\cite{LT:book}.
The striking phenomenon is that the exponential decay rate for different
values of $\alpha$ can be statistically distinguished to allow the Berry
phase of the quasiparticles to be unequivocally detected, leading to the
birth of chaos based Berry phase detectors. We note that in microcavity
optics, classical chaos can be exploited to generate lasing with a high
quality factor and good emission directionality at the same
time~\cite{NSC:1994,MNCSC:1995,NSCGC:1996,NS:1997,GCNNSFSC:1998,NHJS:1999,
WH:2008,Altmann:2009,JSZYWWGLYX:2017,Yang:2018,Bittneretal:2018}.

\section{Hamiltonian and Dirac electron optics} \label{sec:model}

\begin{figure} [hbt!]
\centering
\includegraphics[width=\linewidth]{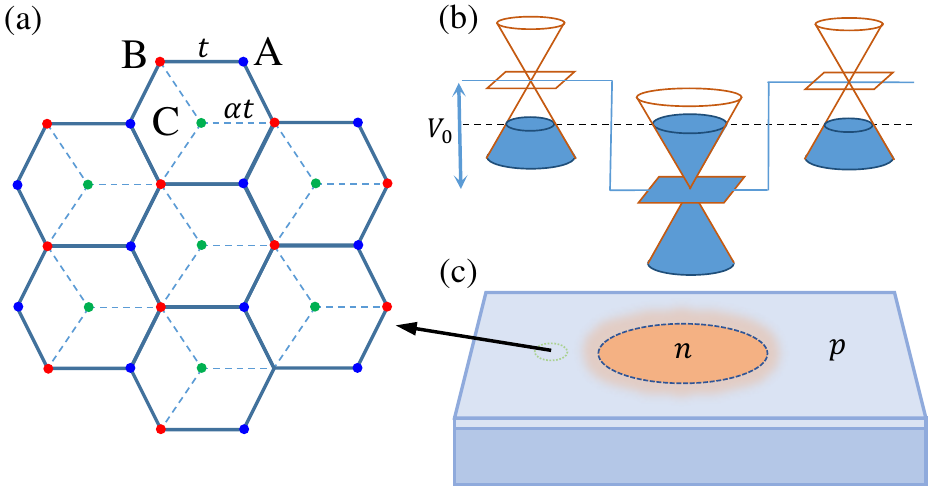}
\caption{ Schematic illustration of an $\alpha$-T$_3$ cavity
and the energy dispersion relation. (a) $\alpha$-T$_3$ lattice structure. 
(b) The electron and hole energy dispersion relations in different spatial
regions. (c) A possible scheme of experimental realization of the cavity
through an applied gate voltage. The amount of the voltage is such that the
quasiparticles are in the Klein-tunneling regime.}
\label{fig:alpha-T3-cavity}
\end{figure}

The $\alpha$-T$_3$ lattice system has the advantage of generating a continuous
spectrum of quasiparticles with systematically varying Berry phase through the
tuning of the value of the parameter $\alpha$ in the unit interval. At the
two opposite ends of the spectrum, i.e., $\alpha = 0,1$, the quasiparticles
are pseudospin-1/2 Dirac fermions and pseudospin-1 Dirac-Weyl particles,
respectively. As illustrated in Fig.~\ref{fig:alpha-T3-cavity}, the lattice
has three nonequivalent atoms in one unit cell, and the interaction strength
is $t$ between $A$ and $B$ atoms and $\alpha t$ between $B$ and $C$ atoms,
where $t$ is the nearest neighbor hopping energy of the graphene lattice. A
cavity of arbitrary shape can be realized by applying an appropriate gate
voltage through the STM technique~\cite{Zhaoetal:2015,Ghaharietal:2017,
Gutierrezetal:2018}, as shown in Fig.~\ref{fig:alpha-T3-cavity}(c). We
consider circular and stadium shaped cavities that exhibit integrable and
chaotic dynamics, respectively, in the classical
limit~\cite{Stockmann:book}. The low-energy Hamiltonian for the
$\alpha$-T$_3$ system about a $K$ point in the hexagonal Brillouin zone
is~\cite{RMFPM:2014,IN:2017} $\hat{H} = \hat{H}_{kin} + V(x)\hat{I}$, where
$\hat{H}_{kin}$ is the kinetic energy, $V(x)$ is the applied potential that
forms the cavity, and $I$ is the $3 \times 3$ identity matrix. The coupling
strength $\alpha$ can be conveniently parameterized as $\alpha = \tan{\psi}$.
The kinetic part of the rescaled Hamiltonian (by $\cos{\psi}$) is
\begin{equation}
\hat{H}_{kin} =
\begin{bmatrix}
    0                 & {f_k \cos \psi}   & 0 \\
    {f^*_k \cos \psi} & 0                 & {f_k \sin \psi} \\
    0                 & {f^*_k \sin \psi} & 0
\end{bmatrix},
\end{equation}
where $f_k = v_F (\xi k_x - ik_y)$, $v_F$ is the Fermi velocity,
$\boldsymbol{k}= (k_x, k_y)$ is the wave vector, and $\xi=\pm$ is the
valley quantum number associated with $K$ and $K'$, respectively. In the
semiclassical regime where the particle wavelength is much smaller than the
size of the cavity so that the classical dynamics are directly relevant,
the DEO paradigm can be instated to treat the particle escape problem,
which is analogous to decay of light rays from a dielectric cavity. In DEO,
the essential quantity is the transmission coefficient of a particle through
a potential step, which can be obtained by wavefunction matching
as~\cite{IN:2017}
\begin{align} \label{eq:alpha-T3-trans}
T = \frac{4ss'\cos{\theta} \cos{\phi}}{2 + 2ss' \cos{(\theta + \phi)}
- \sin^2{2\psi} (s\sin{\theta} - s'\sin{\phi})^2},
\end{align}
where $s=\pm$ and $s'=\pm$ with the plus and minus signs denoting the
conduction and valence band, respectively, and incident and transmitted
angles are $\phi$ and $\theta$, respectively. Imposing conservation of the
component of the momentum tangent to the interface, we get
\begin{displaymath}
\sin{\theta} = (E/|E-V_0|)\sin{\phi}. 
\end{displaymath}
(More details about electron transmission through a potential step can be 
found in Appendix~\ref{Appendix_A}). Our focus is on the survival
probability of the quasiparticles from an $\alpha$-T$_3$ cavity for the
entire material spectrum: $0 \le \alpha \le 1$.

We set the amount of the applied voltage such that the energy range of the
quasiparticles is $V_0/2 < E < V_0$ (the Klein tunneling regime). In the
optical analog, the corresponding relative effective refractive index inside
the cavity is $n = E/(E-V_0)$ and that outside of the cavity is $n=1$. Due to
Klein tunneling, the range of relative refractive index in the cavity is
negative: $-\infty < n < -1$. As a result, a critical angle exists for the
tunneling of electrons through a simple static electrical potential step,
which is $\sin{\phi_c} = (V_0 - E)/E$ and is independent of the $\alpha$
value~\cite{IN:2017}. This behavior is exemplified in the polar representation
of the transmission in Fig.~\ref{fig:SPTD-Intvschaos}(a), which shows that
the value of the transmission increases with $\alpha$. As the value of $\alpha$
is varied in the unit interval, the critical angle remains unchanged.

\section{Results} \label{sec:results}

\subsubsection{Algebraic decay of $\alpha$-T$_3$ quasiparticles from a circular
(integrable) cavity}

The classical phase space contains Kolmogorov-Arnold-Moser (KAM)
tori and an open area through which particles (rays) escape. Initializing
an ensemble of particles (e.g., $10^7$) in the open area, the survival
probability time distribution (SPTD) is given by
\begin{align} \label{eq:SPTD1}
P_{sv}(t) = \int^L_0 ds \int^{p_c}_{-p_c} dp I(s,p)R(p)^{N(t)},
\end{align}
where $L$ is the boundary length, $p_c = \sin{\phi_c} = 1/|n|$ 
with $\phi_c$ being the critical angle for total internal reflection, 
$R(p)=1-T$ is the reflection coefficient for the $\alpha$-T$_3$ quasiparticles 
with transmission $T$ defined in Eq.~(\ref{eq:alpha-T3-trans}), 
$N(t) = t/(2\cos{\phi})$ is the number of bounces off the
boundary, and $I(s,p) = |n|/2L$ is the uniform initial distribution. 

\begin{figure}[htp!]
\centering
\includegraphics[width=\linewidth]{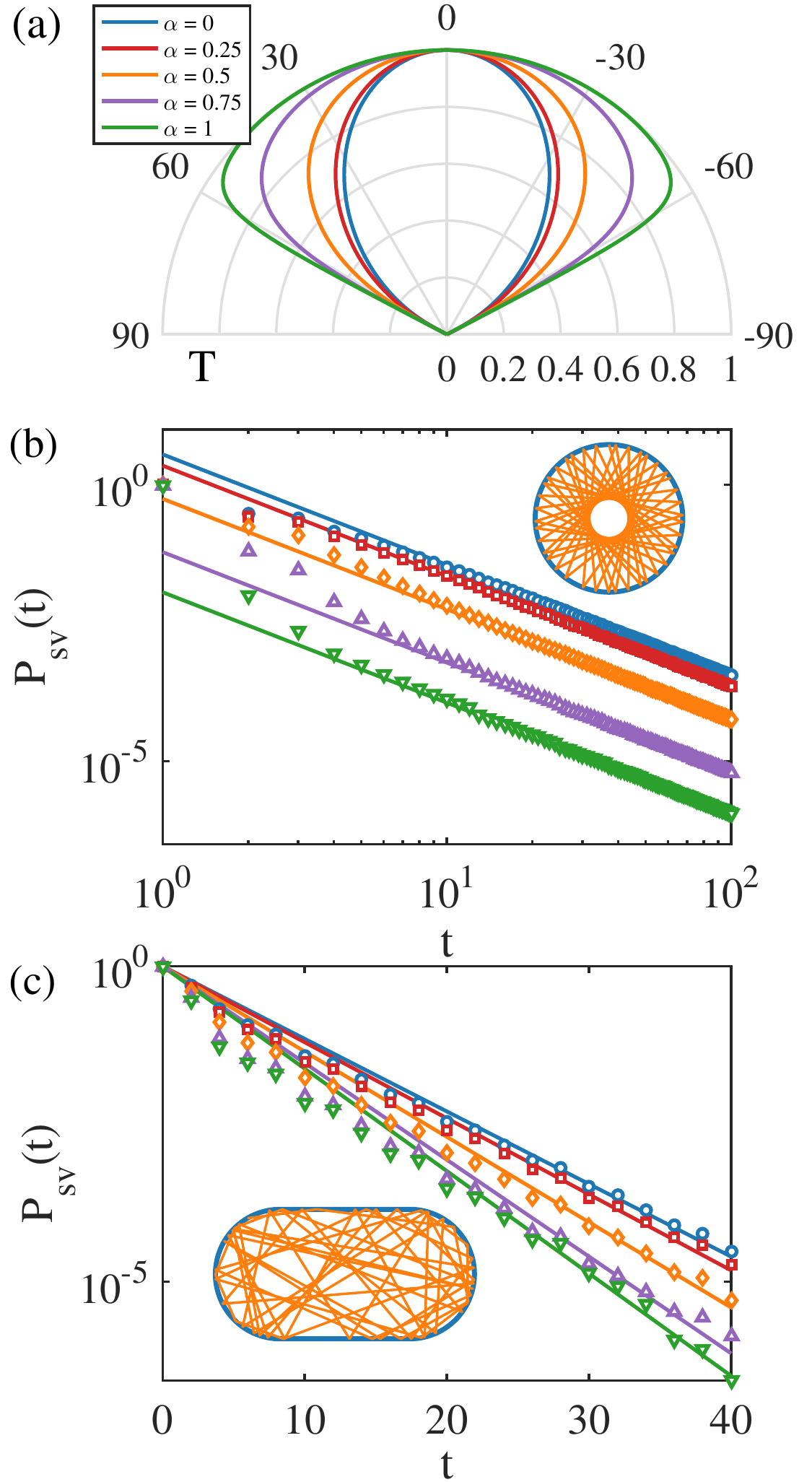}
\caption{ Semiclassical decay of quasiparticles from a cavity in an
$\alpha$-T$_3$ lattice. For particle energy  $E = 0.53 V_0$ (within Klein
tunneling regime) and relative refractive index $n=-1.1277$ inside of the 
cavity, (a) transmission $T$ across a potential step as a 
function of incident angle $\phi$ for a number of equally spaced $\alpha$ 
values. (b) SPTD for the circular (integrable) cavity on a double logarithmic 
plot, where the blue circles, red squares, orange diamonds, purple
up-triangles, and green down-triangles are numerical results for the five
$\alpha$ values in (a), respectively, and the solid lines are the theoretical
predictions. The decay is algebraic but the decay exponent is a constant
independent of the value of $\alpha$. (c) SPTD for a stadium shaped (chaotic)
cavity of semicircle radius one and straight edge of length two	
on a semi-logarithmic plot. The color legends are the same as in (b).
In this case, the decay is exponential and its rate depends on the value of
$\alpha$. Measuring the exponential decay rate then gives the value of
$\alpha$ and the corresponding Berry phase of the underlying material
lattice system.}
\label{fig:SPTD-Intvschaos}
\end{figure}

Consider a circle of unit radius. Using the length of the ray trajectory as
the time scale, we can rewrite Eq.~(\ref{eq:SPTD1}) as
\begin{align} \label{eq:SPTD2}
P_{sv}(t) = |n| \int^{\phi_c}_{0} d\phi \cos{\phi}
\exp{[-\frac{t}{2\cos{\phi}}\ln{(\frac{1}{R})}]},
\end{align}
with
\begin{align}
R^{-1} =  1 + \frac{-4\cos{\theta}\cos{\phi}}{2 + 2\cos{(\theta - \phi)} 
- \sin^2{2\psi}(\sin{\theta} + \sin{\phi})^2}.
\end{align}
The behavior of the particle transmission coefficient shown in 
Fig.~\ref{fig:SPTD-Intvschaos}(a) indicates that particles near the critical 
angle $\phi_c$ can survive for a longer period of time in the cavity.
We can then expand the $\ln(\frac{1}{R})$ term about the critical angle 
$\phi_c$ by defining a new variable $\chi$ with $\phi = \phi_c - \chi$ and 
exploiting the approximation $\chi \to 0$. We have
\begin{align} \label{R-rev}
\ln{(\frac{1}{R})} \approx \frac{4\sqrt{2|n|\cos{\phi_c}}\cos{\phi_c}}
{2 + [2|n|-\sin^2{2\psi}(|n|+1)^2\sin^2{\phi_c}]}\cdot \chi^{1/2}.
\end{align}
Substituting Eq.~(\ref{R-rev}) into Eq.~(\ref{eq:SPTD2}), we obtain the 
SPTD as
\begin{align} \label{eq:algebraic_decay}
P_{sv}&=\frac{1}{4}t^{-2}\{2+[2|n|-\sin^2{2\psi}(|n|+1)^2]\frac{1}{|n|^2}\}^2 \nonumber \\
&= C(n,\psi)t^{-2}
\end{align}
This indicates that the quasiparticles decay algebraically from the cavity 
and the value of the decay exponent is two, regardless of the value of 
$\alpha$. For certain value of $|n|$, as the value of $\alpha$ changes from 
zero to one, the decay coefficient $C(n,\psi)$ decreases, as shown in
Fig.~\ref{fig:SPTD-Intvschaos}(b). Here, SPTD for the circular cavity 
is calculated with $10^7$ random initial points in the open region of the 
phase space. The trajectory from each point is traced with the reflection 
coefficient $R(p)$ at boundary. The survival probability between $t$ and 
$t+\Delta$ with $\Delta = 1$ is calculated with the initial probability one 
at $t=0$. From Fig.~\ref{fig:SPTD-Intvschaos}(b), we see that both theoretical 
and numerical results show an algebraic behavior in the long time regime 
with the exponent of two.

Experimentally, to distinguish the nature of the quasiparticles and to detect
the Berry phase, the decay coefficient is not a desired quantity to measure
as it reflects the short time behavior of the decay process. In fact, it not
only depends the nature of the material (as determined by the value of
$\alpha$) but also on the detailed system design. The long time behavior of
the decay is characterized by the algebraic decay exponent, which does not
depend on the details of the experimental design and, hence, it can possibly
be exploited for Berry phase detection. However, for an integrable cavity,
the algebraic decay exponent remains constant as the value of $\alpha$ is
changed, as shown in Fig.~\ref{fig:SPTD-Intvschaos}(b). It is thus not
feasible to distinguish the quasiparticles by their long time behavior,
ruling out integrable cavities as a potential candidate for detecting the
Berry phase.

\subsubsection{Exponential decay of $\alpha$-T$_3$ quasiparticles from a chaotic
cavity}

\begin{figure} [hbt!]
\centering
\includegraphics[width=0.8\linewidth]{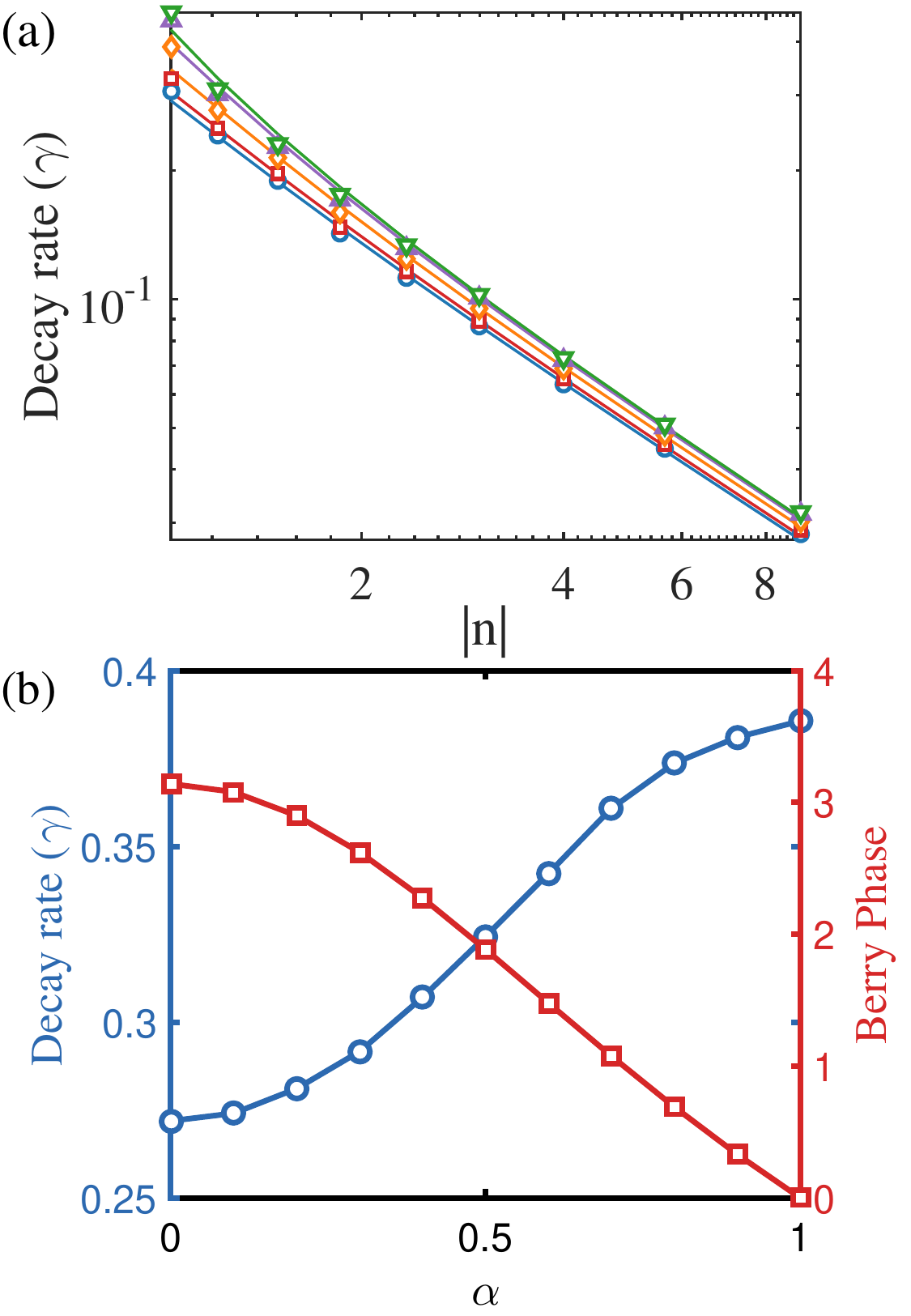}
\caption{ Dependence of the semiclassical exponential decay rate from a
chaotic cavity on the effective refractive index and the detection
of the Berry phase. (a) For $\alpha=0,0.25,0.5,0.75,1$, the decay rate versus
the refractive index, where the blue circles, red squares, orange diamonds,
purple up-triangles and green down-triangles are the respective numerical
results and the dashed curves are theoretical predictions.
(b) For $E/V_0 = 0.53$, detection of Berry phase (red squares) based on
the decay rate (blue circles). As the value of $\alpha$ is changed from
zero to one, there is a one-to-one correspondence between the exponential
decay rate and the Berry phase.}
\label{fig:SPTD-BP}
\end{figure}

For the stadium cavity, the classical dynamics are chaotic, leading to
random changes in the direction of the propagating ray. In this case, the
survival probability of the quasiparticles in the cavity decays exponentially
with time, as shown in Fig.~\ref{fig:SPTD-Intvschaos}(c), where the long time
behavior is determined by the exponential decay rate. The striking phenomenon
is that the decay rate increases monotonically as the value of the material 
parameter $\alpha$ is increased from zero to one, suggesting the possibility 
of using the exponential decay rate to distinguish the $\alpha$-T$_3$ 
materials and to detect the intrinsic Berry phase. The difference in the decay 
rate can be further demonstrated by calculating its dependence on the absolute 
value $|n|$ for different values of $\alpha$, as shown in 
Fig.~\ref{fig:SPTD-BP}(a). For small values of $|n|$, the difference in the 
decay rate is relatively large, indicating a stronger ability to discern the 
$\alpha$-T$_3$ quasiparticles. For large values of $|n|$, the
difference in the decay rate is somewhat reduced. This is expected
because, as the value of $|n|$ is increased from one, the transmission for the
materials at the two ends of the $\alpha$-T$_3$ spectrum, namely graphene and
pseudospin-1 lattice, decreases continuously. For $|n|\rightarrow\infty$,
the transmission tends to zero. This result indicates that, the optimal
regime to discern the quasiparticles for $\alpha$-T$_3$ occurs for $|n|$
above one but not much larger, corresponding to the regime where the particle
energy is slightly above half of the potential height.

In general, for a given value of $\alpha$, the exponential decay rate is
inversely proportional to $n$, which can be argued, as
follows~\cite{RLKP:2006,LRRKCK:2004}. For $P_{sv}(t) \sim \exp{(-\gamma t)}$,
we have $dP_{sv}(t)/dt \sim -\gamma \cdot P_{sv}(t)
\sim -(\langle T(p) \rangle/\langle d \rangle)\cdot P_{sv}(t)$,
where $\langle T(p) \rangle$ and $\langle d \rangle$ are the average
transmission and the distance between two consecutive collisions in the chaotic
cavity. The decay rate can then be obtained in terms of the steady probability
distribution $P_s(s,p)$ as:
\begin{align}
\gamma = \langle T(p) \rangle/\langle d \rangle
= \langle d \rangle^{-1}\int^L_0 ds \int^1_{-1} dp P_s(s,p) T(p) 
\end{align}

In the Klein tunneling regime $V_0/2 < E < V_0$ ($-\infty < n < -1$), we can 
derive an analytical expression for the exponential decay rate based on a 
simple model of the steady probability distribution (SPD) for the 
stadium-shaped cavity that generates fully developed chaos in the
classical limit~\cite{RLKP:2006}. Specifically, we assume that the SPD is a 
uniform distribution over the whole phase space except the open regions 
related to the linear segments of the stadium boundary. The decay rate can 
then be expressed in terms of the steady probability distribution:
\begin{align} \label{Eq:gamma}
\gamma = \frac{2\pi R}{2(\pi A / L)(L - 2l/|n|)}\int^{1/|n|}_{-1/|n|}dp T(p),
\end{align}
where $T(p)$ is the transmission coefficient defined in 
Eq.~(\ref{eq:alpha-T3-trans}), the average path length of ray trajectory 
segments between two successive bounces is $\langle d \rangle = \pi A /L$, 
with $A = \pi R^2 + 2Rl$ and $L = 2\pi R + 2l$ being the area and boundary 
length of the stadium, respectively. Substituting the expressions 
$\sin{\theta} = |n|p$, $\cos{\theta} = -\sqrt{1 - \sin^2\theta} 
= -\sqrt{1 - n^2p^2}$, $\sin{\phi} = p$, and $\cos{\phi} = \sqrt{1 - p^2}$ 
into the expression of $T(p)$, we get
\begin{align} \label{eq:app_trans}
T = 4\sqrt{1-p^2}\sqrt{1 - n^2p^2}/[2 + 2\sqrt{1 - p^2}\sqrt{1 - n^2p^2}
\nonumber \\
+ 2|n|p^2 - \sin^22\psi (n^2p^2 + p^2 + 2|n|p^2)].
\end{align}
In the limit $|n| \approx 1$, imposing change of variable $x = n^2p^2$ to 
get $dp = dx/(2|n|\sqrt{x})$, we can write the decay rate in terms of 
variable $x$ as
\begin{align} \label{Eq:gamma_appox1}
\gamma &= \frac{2\pi R}{2(\pi A / L)(L - 2l/|n|)}\int^{1}_{0}\frac{dx}{\sqrt{x}} T(x) \nonumber \\
&= \frac{2\pi R}{2(\pi A / L)(L - 2l/|n|)}
\int^{1}_{0} \frac{dx}{\sqrt{x}} (1 - x)(1 - \sin^2{2\psi} \cdot x)^{-1}  \nonumber \\
&=\frac{2\pi R}{2(\pi A/L)(L-2l/|n|)}B(1/2,2)F(1,1/2;5/2;\sin^2{2\psi}) \nonumber \\
&\approx \frac{2\pi R}{2(\pi A / L)(L - 2l/|n|)} \frac{4}{3}\cdot
(1+\frac{1}{5}\sin^2{2\psi} + \ldots),
\end{align}
where $B(x,y)=\Gamma(x)\Gamma(y)/\Gamma(x+y)$ is the beta function
and $F(\alpha,\beta;\gamma;z)$ is the Gauss hypergeometric function.

In the $|n|\gg 1$ regime, we use the change of variable $x = np$ to simplify
the decay rate integral. The decay rate becomes
\begin{align} \label{Eq:gamma_appox2}
\gamma = \frac{4\pi R}{2(\pi A / L)L |n|}\int^{1}_{0}
	\frac{4\sqrt{1-x^2}}{2 + 2\sqrt{1 - x^2} - \sin^2 (2\psi) x^2},
\end{align}
which is inversely proportional to the absolute value of the refractive index
$|n|$. More importantly, the decay rate depends on the material parameter 
$\alpha$ monotonically ($\alpha = \tan\psi$, with $\alpha$ increasing from 
zero to one). We note that, the theoretical results in 
Fig.~\ref{fig:SPTD-Intvschaos}(c) is obtained by doing the integration formula
(\ref{Eq:gamma}) directly. The approximation used to derive E
qs.~(\ref{Eq:gamma_appox1}) and (\ref{Eq:gamma_appox2}) is to facilitate 
an analytic demonstration of the scaling of the decay rate with $n$. The 
formulas also reveal that the decay rate increases monotonically with $\alpha$.

Numerically, we choose the stadium shape with the semicircle radius to be one 
and the length of the straight long edge to be two. In the calculation, we 
use a random ensemble of $10^7$ initial points spread over the whole phase 
space and trace the survival probability with time, which is scaled by the 
length of trajectory as in the case of a circular cavity. The numerical 
results are consistent with the theoretical cases based on SPD approximation, 
as shown in Fig.~\ref{fig:SPTD-Intvschaos}(c).

\subsubsection{Detection of Berry phase}

The Berry phase associated with an orbit in the conical bands is given
by~\cite{ICN:2015}
\begin{align}
\phi^B_{\xi} = \pi\xi\cos{(2\psi)}=\pi\xi (\frac{1- \alpha^2}{1 + \alpha^2}).
\end{align}
For the flat band, the Berry phase is
\begin{align}
\phi^B_{0,\xi}=-2\pi\xi\cos{(2\psi)}=-2\pi\xi(\frac{1- \alpha^2}{1+\alpha^2}).
\end{align}
We take $\xi = \pm 1$ for the $K$ and $K'$ valleys, respectively. For
$\xi = 1$, the dependence of the Berry phase on $\alpha$ is shown in
Fig.~\ref{fig:SPTD-BP}(b). As the value of $\alpha$ is increased from zero
to one, the Berry phase decreases monotonically from $\pi$ to zero.
At the same time, the exponential decay rate increases monotonically.
There is then a one-to-one correspondence between the decay rate and the
Berry phase for the entire spectrum of $\alpha$-T$_3$ materials, justifying
a semiclassical chaotic cavity as an effective Berry phase detector.

\section{Discussion} \label{sec:discussion}

To summarize, we uncover a phenomenon in relativistic quantum
chaos that can be exploited to detect the Berry phase of two-dimensional
Dirac materials. In particular, for the spectrum of $\alpha$-T$_3$ materials,
in the semiclassical regime, the decay of the quasiparticles from a chaotic
cavity depends on the intrinsic material parameter. Experimentally, the cavity
can be realized through a gate voltage, where locally the boundary of the
cavity is effectively a potential step. When the Fermi energy of the
quasiparticles is above half but below the potential height, the system is
in the Klein tunneling regime, rendering applicable Dirac electron optics.
In this case, the relative effective refractive index inside the cavity is
between negative infinity and minus one, so a critical angle exists for the
semiclassical ray dynamics. Because of the close interplay between Klein
tunneling and the value of the Berry phase, measuring the quasiparticle
escape rate leads to direct information about the Berry phase and for
differentiating the $\alpha$-T$_3$ materials. Our analysis and calculation
have validated this idea - we have indeed found a one-to-one correspondence
between the exponential decay rate and the value of the Berry phase. In
terms of basic physics, our finding builds up a connection, for the first
time, between classical chaos and Berry phase. From an applied standpoint,
because of the fundamental importance of Berry phase in determining the
quantum behaviors and properties of materials, our work, relative simplicity
notwithstanding, provides an effective and experimentally feasible way to
assess the Berry phase for accurate characterization of the underlying
material. This may find broad applications in materials science and engineering
where new nanomaterials are being discovered at a rapid pace, demanding
effective techniques of characterization.

A possible experimental scheme to detect the Berry phase for the family of
$\alpha$-T$_3$ materials is as follows. For each type of material, one first
makes a chaotic cavity (e.g., a stadium or a heart shaped domain). One then
measures the quasiparticle decay rate for the graphene cavity (corresponding
to $\alpha = 0$). Since the Berry phase of graphene is known, one can use
the measurement as a baseline for calibrating the results from other materials 
in the family. Finally, making use of the one-to-one correspondence between the 
curves of the decay rate and the Berry phase versus the material parameter
$\alpha$ as theorized in this paper, one can detect the actual Berry phase
for the material with any value of $\alpha$ for $0 < \alpha \le 1$.

\section*{Acknowledgment}

We would like to acknowledge support from the Vannevar Bush Faculty Fellowship 
program sponsored by the Basic Research Office of the Assistant Secretary of 
Defense for Research and Engineering and funded by the Office of Naval 
Research through Grant No.~N00014-16-1-2828.

\appendix

\section{Band structure and wavevectors across a potential step}
\label{Appendix_A}

\begin{figure*}
\centering
\includegraphics[width=\linewidth]{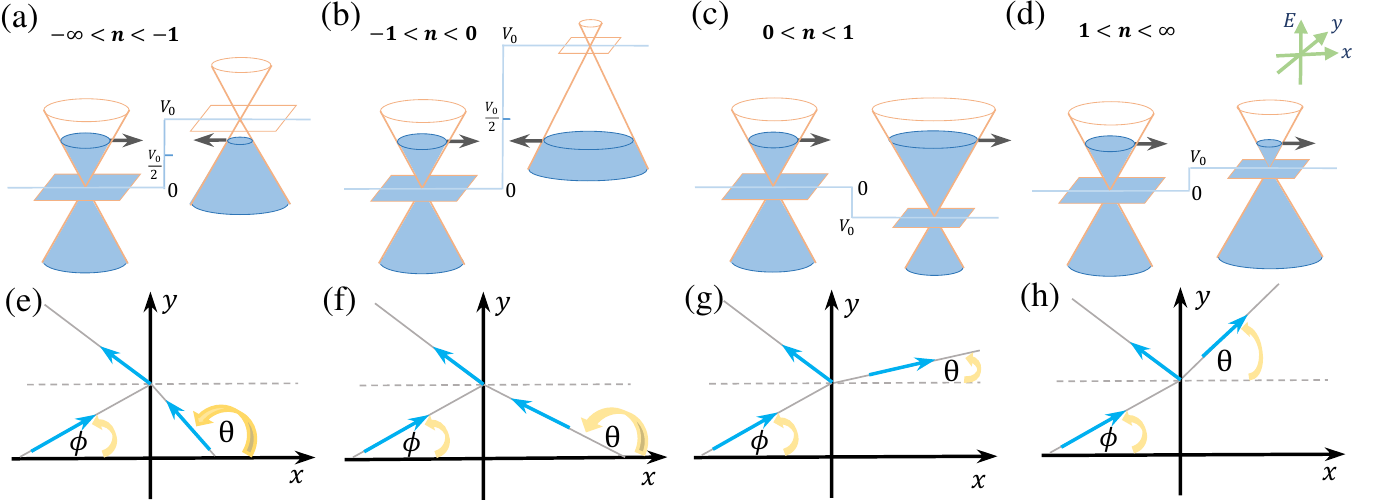}
\caption{Schematic illustration of the band structures 
and wavevectors across a potential step with different values of the 
refractive index. First row: band structures across the potential step 
with different values of the gate potential (corresponding to different
values of the refractive index) at fixed Fermi energy. The black arrows 
denote the wavevector directions (only the cases with the wavevector in 
the $x$ direction are shown). In the regime of negative refractive index,
the wavevector directions are reversed. Second row: electron wavevectors 
with the incident and transmitted angles $\phi$ and $\theta$, respectively.}
\label{fig:SS1}
\end{figure*}

To better understand the optical-like decay behavior of quasiparticles from 
a cavity formed by an electrostatic gate potential, we illustrate the 
electron band structure and the wavevectors across a potential step 
associated with a transmission process, as shown in Fig.~\ref{fig:SS1}. 
We also indicate a classification scheme of the regimes with different 
values of the refractive index, which are determined by different values
of the applied potential relative to the Fermi energy. In particular,
there are regimes of positive and negative values of the refractive index
with respect to cases where a critical angle exists or is absent. For
convenience, the incident electron is assumed to be in the conduction band, 
i.e., with a positive Fermi energy, and we vary the potential height $V_0$. 
When $V_0$ is larger than the Fermi energy, the transmitted electron is in 
the valence band. In this case, the wavevector has a negative $x$ and a 
positive $y$ component but the direction of the velocity remains unchanged, 
leading to a negative value of the refractive index.

More specifically, for gate potential height in the range $V_0/2<E<V_0$, the 
value of the refractive index $n = E/(E-V_0)$ falls in the range 
$-\infty < n < -1$. There is a critical angle in this case, which is 
determined by $\sin{\theta} = 1 = (E/|E-V_0|)\sin{\phi_c}$. The 
transmission angle can be obtained in terms of incident angle $\phi$ as
\begin{align} \label{Eq:angle-relation_1}
\theta &= \pi - \tan^{-1}\frac{\sin \phi \cdot E/V_0}{\sqrt{(1-E/V_0)^2 
	- (\sin\phi \cdot E/V_0)^2}} \nonumber \\
& = \pi + \tan^{-1}\frac{n \sin\phi}{\sqrt{1 - (n\sin\phi)^2}}.
\end{align}
where the relations $\sin{\theta} = (E/|E - V_0|)\sin{\phi}$ and 
$\cos{\theta} = -\sqrt{1 - \sin^2{\theta}}$ have been used. The band 
structure and angles corresponding to the wavevectors are shown in 
Figs.~\ref{fig:SS1}(a,e), respectively.

In the regime where the potential height satisfies $0<E<V_0/2$, the value
of the refractive index is in the range $-1<n<0$. As a result, there is 
no critical angle. The transmission angle can be obtained in the same form 
as Eq.~(\ref{Eq:angle-relation_1}). A schematic illustration of the band 
structure and the wavevector angles for this case are shown in
Figs.~\ref{fig:SS1}(b,f), respectively.

For $V_0<0<E$, the value of the refractive index is in the positive range 
$0<n<1$, because both the incident and transmitted electron is in the 
conduction band. There is no critical angle in this case. The transmission 
angle can be obtained as
\begin{align} \label{Eq:angle-relation_2}
\theta &= \tan^{-1} \frac{\sin\phi \cdot E/V_0}{\sqrt{(1-E/V_0)^2 - (\sin\phi\cdot E/V_0)^2}} \nonumber \\
&= \tan^{-1}\frac{n\sin\phi}{1 - (n\sin\phi)^2}.
\end{align}
where the relations $\sin{\theta} = [E/(E-V_0)]\sin{\phi}$ and 
$\cos{\theta} = \sqrt{1 - \sin^2{\theta}}$ are used. The band structure 
and wavevectors related angles are depicted in Figs.~\ref{fig:SS1}(c,g),
respectively.

In the regime $0<V_0 <E$, the refractive index is in the range $1<n<\infty$ 
with both the incident and transmitted electron in the conduction band. 
There is a critical angle in this case determined by 
$\sin{\theta} = 1 = [E/(E-V_0)]\sin{\phi_c}$. The transmission angle 
can be obtained in the same form as in Eq.~(\ref{Eq:angle-relation_2}). The
band structures and wavevectors are illustrated in Figs.~\ref{fig:SS1}(d,h),
respectively.

\section{Survival probability distribution of $\alpha$-T$_3$ quasiparticles in
different energy regimes}

For completeness, we derive the decay law of the survival probability of
$\alpha$-T$_3$ quasiparticles and obtain the decay rate in other energy regimes
than the Klein tunneling regime. We argue that the decay law in these regimes
is practically infeasible for detecting the Berry phase. For example, in the
regimes where there is no critical angle, the decay can be too fast for it 
to be useful. In the regimes where there is a critical angle, the decay for 
distinct quasiparticles from the material family follows a similar law, making 
it difficult to distinguish the different quasiparticles. 

\subsection{The $0< E < V_0/2$ regime}

\begin{figure*}
\centering
\includegraphics[width=\linewidth]{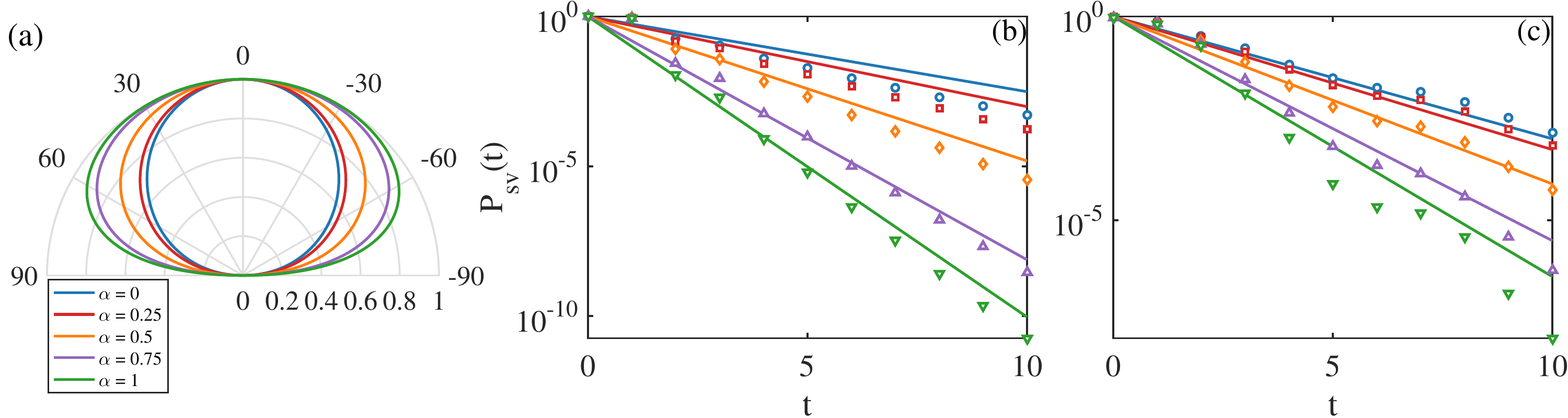}
\caption{ Survival probabilities from integrable and chaotic cavities
for $0< E < V_0/2$. For $E/V_0 = 1/3$ and $n=-0.5$, (a) transmission versus
the incident angle on a polar plot, (b) decay of the survival probability
from a circular (integrable) cavity with time, and (c) decay of the survival
probability from a stadium shaped (chaotic) cavity.}
\label{fig:SPTD-Exp}
\end{figure*}

In this energy regime, the refractive index $n=E/(E-V_0)$ of the cavity is in the
range $-1<n<0$. In this regime, there exists no critical angle for rays inside
the cavity. Figure~\ref{fig:SPTD-Exp}(a) shows that the transmission is
nonzero for all angles and it increases with decreasing $\alpha$ values. In
this case, the decay of quasiparticles is exponential and it does not depend
on the nature of the classical dynamics, i.e., integrable or chaotic, as
shown in Fig.~\ref{fig:SPTD-Exp}. 

A theoretical explanation of the features in Fig.~\ref{fig:SPTD-Exp} is as
follows. Due to the absence of a critical angle for Dirac electron optical
rays in the energy range $0 < E < V_0/2$, the survival probability from a
circular (integrable) is mainly determined by the ray behavior about
$\phi=\pi/2$. Letting $\phi = \pi/2 - x$, where $x$ is a small angle deviation
from $\pi/2$, and using the approximations
\begin{align}
&\sin\phi \approx \sin \phi_c - \cos\phi_c\cdot x, \nonumber \\
&\cos \phi \approx \cos \phi_c + \sin\phi_c\cdot x, \nonumber \\
&\sin \theta \approx |n|\cdot (\sin\phi_c - \cos\phi_c\cdot x), \nonumber \\
&\cos \theta \approx -\sqrt{1 - n^2 \cdot (\sin\phi_c -\cos\phi_c \cdot x)^2}, \nonumber
\end{align}
we get
\begin{align}
\ln R^{-1} = \frac{4 x\sqrt{1-n^2}}{2+2|n| - \sin^2{(2\psi)} (1 + |n|)^2}.
\end{align}
where $R = 1 - T$ with $T$ being the transmission coefficient defined in Eq.~(\ref{eq:alpha-T3-trans}) in the main text. The survival probability can be expressed as
\begin{align}
P_{sv}=\exp{\{-\frac{2\sqrt{1-n^2}}{2+2|n|-\sin^2{(2\psi)}(1+|n|)^2}\cdot t\}}
\end{align}

For a chaotic cavity, the angle distribution is random, leading to an
exponential behavior of the survival probability. We can obtain the
expression for the decay rate $\gamma$ by approximating $P_{sv}$ as
\begin{equation}
P_{sv}(t)\approx\langle1-T(p)\rangle^{t/\langle d \rangle}
= \exp{\{\ln{[1-\langle T(p)\rangle]}(t/\langle d \rangle)\}}.
\end{equation}
The decay rate can be expressed as
\begin{equation}
\gamma = -\frac{1}{\langle d \rangle}\ln{[1-\langle T(p)\rangle]}.
\end{equation}	
For either the integrable
or the chaotic cavity, the exponential decay rate depends on the material
parameter $\alpha$ which, in principle, can be used to detect the Berry
phase. However, due to the lack of a critical angle in this energy range,
experimentally it would be difficult to confine the quasiparticles. Indeed,
comparing with the exponential decay from a chaotic cavity in the Klein
tunneling regime ($V_0/2 < E < V_0$) as treated in the main text, here the
decay is much faster.

\subsection{The $0 < V_0 < E$ regime}

\begin{figure*}
\centering
\includegraphics[width=\linewidth]{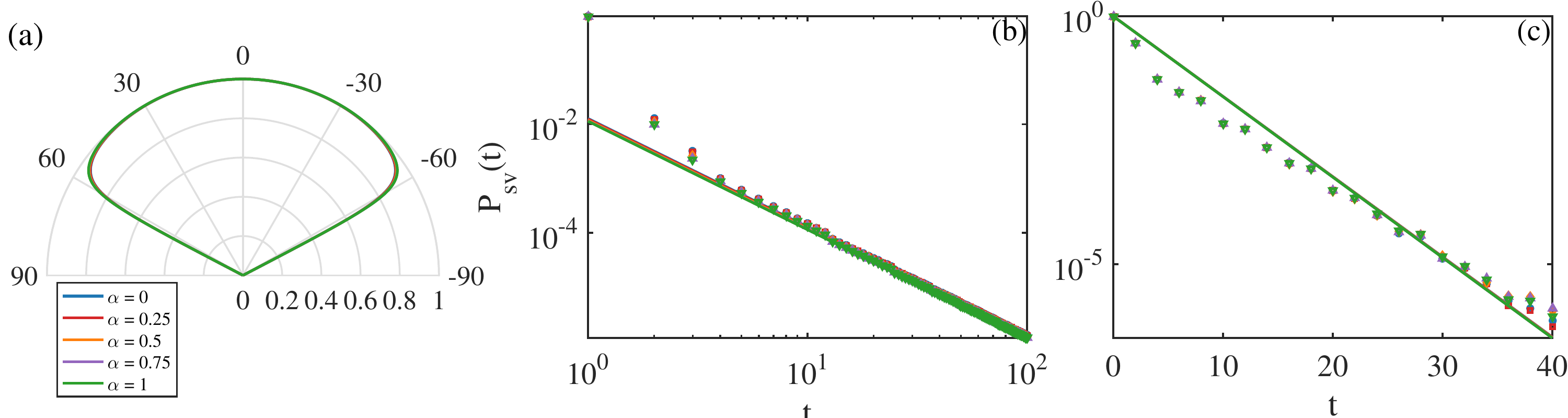}
\caption{ Survival probabilities from integrable and chaotic cavities for
$0 < V_0 < E$. For $E/V_0 = 8.8309$ ($n = 1.1277$), (a) polar representation of
the transmission with respect to the incident angle, (b) decay with time of
the survival probability from a circular (integrable) cavity, and (c) decay of
survival probability from a stadium shaped (chaotic) cavity.}
\label{fig:SPTD-fig6}
\end{figure*}

For the energy range $0 < V_0 < E$ with the refractive index $n=E/(E-V_0)$
of the cavity in the range $1<n<\infty$, the survival probability with time
exhibits an algebraic decay from an integrable cavity and an exponential decay
from a chaotic cavity, which is characteristically similar to the decay
behaviors in the Klein tunneling regime ($V_0/2 < E < V_0$) treated in the
main text. A difference is that, for $0 < V_0 < E$, the dependence of the
transmission on the material parameter $\alpha$ is much weaker in the sense
that, as the value of $\alpha$ is increased from zero to one, the transmission
barely changes. It is thus practically difficult to distinguish the
quasiparticles for different materials. These behaviors are shown in
Fig.~\ref{fig:SPTD-fig6}, where the analytical fitting is calculated
in the same way as in the main text.

\subsection{The $V_0<0<E$ regime}

\begin{figure*}[htp]
\centering
\includegraphics[width=\linewidth]{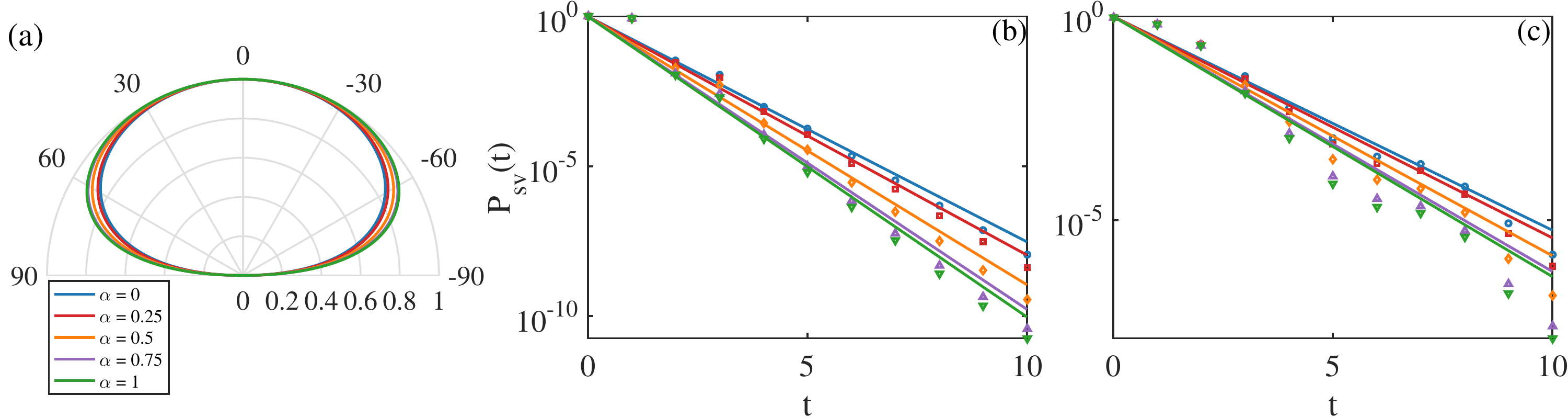}
\caption{ Survival probability from integrable and chaotic cavities
in the $V_0<0<E$ energy regime. For $E/V_0 = -1$ ($n = 0.5$),
(a) a polar representation of the transmission versus the incident angle,
(b) decay of survival probability from a circular (integrable) cavity, and
(c) decay of survival probability from a stadium shaped (chaotic) cavity.}
\label{fig:SPTD-fig7}
\end{figure*}

In the energy regime $V_0<0<E$ with the refractive index $n=E/(E-V_0)$ of the
cavity in the range $0<n<1$, the decay of the survival probability is
similar to that in the $0<E<V_0/2$ regime. In particular, regardless of the
nature of the classical dynamics (integrable or chaotic), the survival
probability exhibits an exponential decay with time, as shown in
Fig.~\ref{fig:SPTD-fig7}. Again, comparing with the energy regime of Klein
tunneling, the decay is much faster here, making experimental detection of
Berry phase difficult.

\section{Comparison between the decay of survival probability for
pseudospin-1/2 and pseudospin-1 quasiparticles}

The best studied material in the $\alpha$-T$_3$ family is graphene,
corresponding to $\alpha = 0$. There is also a growing interest in the
material at the other end of the spectrum: $\alpha = 1$ for which the
quasiparticles are of the pseudospin-1 nature. We offer a comparison of
the decay behavior of the quasiparticles at these two extreme cases.

In the energy range $0 < E < V_0/2$ [corresponding to negative refractive
index: $-1 < n = E/(E-V_0) < 0$], there is no critical angle for total internal
reflection. For both integrable and chaotic cavities, the survival probability
decays exponentially with time, with no qualitative difference. As the
absolute value of the refractive index is increased, the range of angle for
transmission is large for pseudospin-1 quasiparticles, but the range is
smaller for pseudospin-1/2 quasiparticles. For integrable cavities, the
difference is somewhat larger.

In the energy range for Klein tunneling: $V_0/2<E<V_0$ ($-\infty < n< -1$),
a critical angle arises, above which there are total internal reflections.
For an integrable cavity, the survival probability decays algebraically with
time, but the decay is exponential for a chaotic cavity. In the integrable
case, the algebraic decay exponents have approximately identical values for
the pseudospin-1 and pseudospin-1/2 particles. However, for a chaotic cavity,
the decay of pseudospin-1 quasiparticles is much faster than that of
pseudospin-1/2 quasiparticles. Chaos can thus be effective in detecting the
Berry phase to distinguish the two types of quasiparticles. In fact, as
demonstrated in the main text, chaos in the Klein tunneling regime can be
effective for detecting the Berry phase across the entire material spectrum
of the $\alpha$-T$_3$ family.

In the energy range of $V_0<E$ ($1<n<\infty$), a critical angle exists. The
decay behavior of the survival probability is algebraic for an integral
cavity and exponential for a chaotic cavity. The difference in the transmission
versus the incident angle is small for pseudospin-1 and pseudospin-1/2
quasiparticles, leading to a similar value of the algebraic decay coefficient 
in the integrable case and a similar exponential decay law in the chaotic
case. In this energy range, to use the decay behavior to discern the
quasiparticles would be practically difficult.

In the energy range $V_0 < 0 < E$ ($0<n<1$), there is no critical angle, and the
decay behavior is exponential for both integrable and chaotic cavities. As
the energy is increased, the difference in the decay behaviors of pseudospin-1
and pseudospin-1/2 quasiparticles diminishes, ruling out the possibility of
exploiting the decay for detection of Berry phase.

Finally, we note a symmetry related phenomenon: for spin-1 quasiparticles the
behavior of the survival probability is identical for positive and negative
refractive index regimes, as a result of symmetry in the expression of the
transmission coefficient. 


%
\end{document}